\documentclass[12pt]{article}
\usepackage{amsmath}
\def\be{\begin{equation}}
\def\ee{\end{equation}}
\def\arr{\begin{array}{rll}}
\def\ea{\end{array}}
\def\bea{\begin{eqnarray}}
\def\eea{\end{eqnarray}}
\def\N2{$N{=}2$}

\def\>{\rangle}
\def\<{\langle}
\def\+{\dagger}
\def\={\ =\ }

\begin{document}
%\large
\renewcommand{\thefootnote}{\fnsymbol{footnote}}
\begin{titlepage}
\setcounter{page}{0}
\vskip 1cm
\begin{center}
{\LARGE\bf  Self--dual metrics  with maximally}\\
\vskip 0.5cm
{\LARGE\bf    superintegrable geodesic flows}\\
\vskip 2cm
$
\textrm{\Large Sergei Filyukov \ } \textrm{\Large and \ }
\textrm{\Large Anton Galajinsky\ }
$
\vskip 0.7cm
{\it
Laboratory of Mathematical Physics, Tomsk Polytechnic University, \\
634050 Tomsk, Lenin Ave. 30, Russian Federation} \\
\vskip 0.4cm
{E-mails: filserge@tpu.ru,~ galajin@tpu.ru}

\vskip 0.5cm

\end{center}
\vskip 1cm
\begin{abstract} \noindent
A class of self--dual and geodesically complete spacetimes with maximally superintegrable geodesic flows is constructed by applying the Eisenhart lift to mechanics in pseudo--Euclidean spacetime of signature $(1,1)$.
It is characterized by the presence of a second rank Killing tensor. Spacetimes of the ultrahyperbolic signature $(2,q)$ with $q > 2$, which admit a second rank Killing tensor and possess superintegrable geodesic flows, are built.
\end{abstract}

\vspace{0.5cm}

\end{titlepage}

\renewcommand{\thefootnote}{\arabic{footnote}}
\setcounter{footnote}0

\noindent
{\bf 1. Introduction}\\

\noindent

Recent upsurge of interest in Lorentzian spacetimes admitting Killing tensors of rank greater than two \cite{GHKW}--\cite{MC} was triggered by the work in \cite{GHKW}, where the Eisenhart lift \cite{Eis}
was applied to Goryachev--Chaplygin and Kovalevskaya's tops so as to produce irreducible rank--3 and rank--4 Killing tensors. The Eisenhart lift is a variant of geometric description of classical mechanics in which the equations of motion following from some Lagrangian function $\mathcal{L}$ are embedded into the null geodesic equations associated with the Lorentzian metric
$d \tau^2=2 \mathcal{L} dt^2+2 dt ds$, where $s$ is an extra coordinate. A mechanical system with $n$ degrees of freedom thus gives rise to an $(n+2)$--dimensional Lorentzian spacetime. Within such a geometric framework, each integral of motion of the original dynamical system which is a polynomial in momenta turns into a Killing tensor of the metric whose rank is equal to the degree of the polynomial. In particular, in Refs. \cite{GHKW}--\cite{MC} various examples of such a kind have been studied. Irreducible Killing tensors of rank up to $n$ in an $(n+2)$--dimensional spacetime have been built \cite{G}.

The Eisenhart lift yields a Ricci--flat spacetime provided the potential governing the dynamical system is a harmonic function. Unfortunately, none of the potentials compatible with integrals of motion cubic (or higher) in velocities known to date is described by a harmonic function. In particular, none of the Lorentzian metrics proposed in \cite{GHKW}--\cite{MC} solves the vacuum Einstein equations.

The first examples of Ricci--flat spacetimes admitting irreducible higher rank Killing tensors have been constructed in a recent work \cite{CG} by switching from Lorentzian signature to
ultrahyperbolic one. The principal observation in \cite{CG} is that, if one starts with a dynamical system in a pseudo--Euclidean spacetime of signature $(p,q)$, the uplifted model is formulated in a spacetime of the ultrahyperbolic signature $(p+1,q+1)$. Most importantly, the signature also alters the condition on the potential which singles out Ricci--flat spacetimes. In particular, Ricci--flat spacetimes of signature $(2,q)$ with $q=2,3,4$ which admit irreducible rank--3 or rank--4 Killing tensors were constructed by applying the Eisenhart lift to integrable models introduced by Drach \cite{Drach}. Note, however, that the spacetimes in \cite{CG} are not geodesically complete.

In the general relativistic context, Killing tensors are used to establish integrability of geodesic equations. They are indispensable for separating variables in the Hamilton--Jacobi and Sch\"odinger equations as well as
in field theory equations such as the Klein--Gordon and Dirac equations formulated on curved backgrounds. The celebrated examples include the second rank Killing tensor in Kerr geometry \cite{C}--\cite{WP} and its higher--dimensional generalizations \cite{FK,KKPF}.

In this work, we continue the research initiated in \cite{CG} and focus on the case of Killing tensors of valence two. Our principal objective is to study whether geodesically complete spacetimes of
signature $(2,q)$ with $q\geq2$ admitting a second rank Killing tensor can be constructed within the Eisenhart framework.
Below we build a class of self--dual and geodesically complete spacetimes with maximally superintegrable geodesic flows. Spacetimes of the ultrahyperbolic signature $(2,q)$ with $q > 2$, which admit a second rank Killing tensor and possess superintegrable geodesic flows, are discussed as well.

The work is organized as follows. In the next section, we consider a special class of mechanical systems in $d=2$ pseudo--Euclidean spacetime of signature $(1,1)$ whose potential is an additive function $U(x,y)=v(y)+b x+a$, where $v(y)$ is a continuously differentiable function and $a$, $b$ are constants. Within the Eisenhart lift, such potentials give rise to self--dual metrics \cite{CG}. We then show that such systems always admit an extra integral of motion quadratic in momenta and for $b=0$ they are maximally superintegrable\footnote{Recall that a Hamiltonian system with $2n$
phase space degrees of freedom is called Liouville integrable if it admits $n$ functionally
independent integrals of motion $F_i$, $i=1,\dots,n$, which are in involution $\{F_i,F_j\}=0$. If there are more than $n$ such integrals,
the model is called superintegrable. A maximal number of functionally independent integrals
of motion is $2n-1$. Systems possessing $2n-1$ first integrals are called maximally superintegrable. For such models the general solution to the equations of motion can in principle be obtained from the integrals of motion by purely algebraic means.}. In Sect. 3, self--dual metrics with maximally superintegrable geodesic flows are constructed by applying the Eisenhart lift to mechanical systems built in Sect. 2. Spacetimes of the ultrahyperbolic signature $(2,q)$ with $q > 2$, which admit a second rank Killing tensor and possess superintegrable geodesic flows, are discussed in Sect. 4. In the concluding Sect. 5 we summarize our results and discuss possible further developments.
\vspace{0.5cm}

\noindent
{\bf 2. Integrable systems in spacetime of signature $(1,1)$}\\

\noindent
Let us consider a dynamical system in $d=2$ pseudo--Euclidean spacetime of signature $(1,1)$ which is governed by the Hamiltonian
\be\label{H}
H=p_x p_y+v(y)+bx+a.
\ee
Here $(p_x,p_y)$ are momenta canonically conjugate to the configuration space variables $(x,y)$ obeying the conventional Poisson brackets $\{x,p_x\}=1$, $\{y,p_y\}=1$, $v(y)$ is a continuously differentiable function of its argument, and $a$, $b$ are constants.
As was shown in Ref. \cite{CG}, within the Eisenhart framework such models give rise to self--dual solutions of the vacuum Einstein equations formulated in a four--dimensional spacetime of signature $(2,2)$.

Let us study under which circumstances the dynamical system (\ref{H}) admits an extra integral of motion quadratic in momenta. Considering a quadratic polynomial in momenta
\be
I_2=A(x,y) p_x^2+B(x,y) p_y^2+C(x,y) p_x p_y+D(x,y),
\ee
where $A(x,y)$, $B(x,y)$, $C(x,y)$, $D(x,y)$ are functions to be determined, and demanding that it be conserved in time, one obtains the system of linear partial differential equations on the coefficients
\bea\label{se}
&&
\partial_y A(x,y)=0, ~  \partial_x A(x,y)+\partial_y C(x,y)=0,
\\[2pt]\label{e1}
&&
\partial_x B(x,y)=0, ~ \partial_y B(x,y)+\partial_x C(x,y)=0,
\\[2pt]\label{e2}
&&
b C(x,y)+2 B(x,y) v'(y)-\partial_x D(x,y)=0,
\\[2pt]\label{e3}
&&
2 b A(x,y)+C(x,y) v'(y)-\partial_y D(x,y)=0.
\eea
Above we abbreviated $\partial_x=\frac{\partial}{\partial x}$, $\partial_y=\frac{\partial}{\partial y}$ and $v'(y)=\frac{d v(y)}{d y}$. The general solution to the equations (\ref{se}) and (\ref{e1}) reads
\be
A(x,y)=\lambda-\gamma x-\frac{\rho x^2}{2}, \qquad B(x,y)=\sigma-\beta y-\frac{\rho y^2}{2}, \qquad C(x,y)=\alpha+\beta x+\gamma y+\rho x y,
\ee
where $\alpha$, $\beta$, $\gamma$, $\lambda$, $\sigma$ and $\rho$ are constants.

Suppose that both $\partial_x D(x,y)$ and $\partial_y D(x,y)$ are nonzero. Then differentiating Eqs. (\ref{e2}) and (\ref{e3}) with respect to $y$ and $x$, respectively, and subtracting one from another, one obtains the 
restriction $b \rho=0$ which means\footnote{The case $b=0$ is treated below in Eq. (\ref{b0}).}
\be
\rho=0
\ee
and the ordinary differential equation to fix $v(y)$
\be\label{sE}
2(\sigma-\beta y) v''(y)-3 \beta v'(y)+3 b \gamma=0.
\ee
It is straightforward to verify that for $\beta\ne 0$ a solution of (\ref{sE}) is not continuously differentiable everywhere so we discard it, while for $\beta=0$ Eq. (\ref{sE}) yields a quadratic potential\footnote{To be more precise, the potential is a polynomial of the second degree in $y$ which can be brought to the form (\ref{qp}) by a canonical transformation.}
\be\label{qp}
v(y)=g y^2,
\ee
where $g$ is a constant. The corresponding model possesses two quadratic integrals of motion
\bea\label{I1}
&&
I_1=p_x^2+2 b y,
\\[2pt]\label{I2}
&&
I_2=p_x (x p_x-y p_y)+\frac{3 b}{4 g} p_y^2+2 b x y -\frac{2 g}{3} y^3,
\eea
which along with the Hamiltonian (\ref{H}) form the functionally independent set. This case is maximally superintegrable.
It is important to stress that the function (\ref{I1}) Poisson commutes with (\ref{H}). Thus for arbitrarily chosen $v(y)$ the two--dimensional model (\ref{H}) is integrable.

Further specification occurs if one of the partial derivatives $\partial_x D(x,y)$ or $\partial_y D(x,y)$ is allowed to vanish. If $\partial_y D(x,y)=0$, Eq. (\ref{e3}) specifies $v(y)$ to be a linear function of its argument, which is not particularly interesting. Assuming $\partial_x D(x,y)=0$, one reveals two options. Either $\rho=\beta=0$ which leads to a quadratic potential (\ref{qp}) considered above, or $b=0$ which constraints the original potential in (\ref{H}) and results in the maximally superintegrable system
\be\label{b0}
H=p_x p_y+v(y), \qquad I_1=p_x, \qquad I_2=p_x (x p_x-y p_y)-y v(y)+\int\limits_0^y v(r) dr.
\ee

We thus conclude that for arbitrarily chosen parameter $b$ and the potential $v(y)$ the model (\ref{H}) is integrable with the second quadratic invariant exposed in Eq. (\ref{I1}) above, while for $b=0$ it is maximally superintegrable and is described by (\ref{b0}).

\vspace{0.5cm}

\noindent
{\bf 3. Self--dual metrics with maximally superintegrable geodesic flows}\\

\noindent
Superintegrable models of the type (\ref{b0}) offer a valid starting point for the construction of self--dual metrics with maximally superintegrable geodesic flows. It suffices to use the Eisenhart lift adopted to mechanics in $d=2$ pseudo--Euclidean spacetime of signature $(1,1)$.

As the first step, one considers a four--dimensional spacetime of signature $(2,2)$ endowed with the metric
\be\label{metric}
d \tau^2=g_{AB}(z) d z^A d z^B=-2 U(x,y) d t^2+2 dt ds+2 dx dy,
\ee
where $z^A=(t,s,x,y)$ and $U(x,y)$ is an arbitrary function. If $U(x,y)$ is additive
\be\label{cond}
U(x,y)=u(x)+v(y),
\ee
the spacetime is Ricci--flat, while the linear first entry
\be
u(x)=a+b x
\ee
along with arbitrary $v(y)$ provide a self--dual solution \cite{CG}. Note that metrics similar to (\ref{metric}) have been considered by Plebanski \cite{Pl} in the context of the second heavenly equation. They differ from (\ref{metric}) by the dependence of $U$ on its arguments which, in our notation, would be $U(t,y)$.

Null geodesic equations associated with the metric (\ref{metric})
\be\label{geod}
\frac{d^2 z^A}{d \tau^2}+\Gamma^A_{BC} (z) \frac{d z^B}{d \tau} \frac{d z^C}{d \tau}=0, \qquad g_{AB}(z) \frac{d z^A}{d \tau} \frac{d z^B}{d \tau}=0
\ee
are central to the Eisenhart lift of the Hamiltonian dynamics governed by the potential $U(x,y)$
\be\label{ds}
H=p_x p_y+U(x,y).
\ee
Being rewritten in components, Eqs. (\ref{geod}) reproduce the equations of motion which follow form (\ref{ds})
\be\label{em}
\frac{d^2 x}{d t^2}=-\partial_y U, \qquad \frac{d^2 y}{dt^2}=-\partial_x U
\ee
along with
\be\label{first}
\frac{d t}{d \tau}=c_1, \qquad \frac{d s}{d t}-2 U(x,y)=c_2, \qquad \frac{dx}{dt} \frac{dy}{dt}+\frac{ds}{dt}-U(x,y)=0,
\ee
where $c_1$ and $c_2$ are constants of integration. The leftmost relation in (\ref{first}) states that $t$ is affinely related to $\tau$
\be\label{afr}
t=c_1 \tau+t_0,
\ee
while the rightmost equation implies that $-c_2$ should be interpreted as the value of the conserved energy $\frac{dx}{dt} \frac{dy}{dt}+U(x,y)$. The original dynamics is recovered by implementing the null reduction with respect to the variable $s$ \cite{Eis}.

Within the Eisenhart framework, a conserved quantity characterizing the dynamical system (\ref{ds}) which is a polynomial in momenta turns into a Killing tensor of the metric (\ref{metric}) whose rank is equal to the degree of the polynomial. In particular, the leftmost equation in (\ref{first})
and the relations which link the momenta and velocities
\be\label{mv}
p_x=\frac{d y}{d t}, \qquad p_y=\frac{dx}{dt}
\ee
imply that a multiplication of a conserved charge of degree $l$ in momenta by ${\left(\frac{d t}{d \tau}\right)}^l$ yields an expression of the form
$K_{A_1 \dots A_l} (z) \frac{d z^{A_1}}{d \tau}\dots \frac{d z^{A_l}}{d \tau}$ from which the Killing tensor $K_{A_1 \dots A_l} (z)$ is obtained.
Notice that, as compared to the conventional Newtonian mechanics, the momenta and velocities are interchanged in (\ref{mv}). This is to be remembered when obtaining Killing tensors in explicit form.

In this work, we are primarily concerned with the construction of self--dual metrics with maximally superintegrable geodesic flows. To this end, in what follows we set $u(x)=0$ in Eq. (\ref{cond}) and consider metrics specified by a continuously differentiable function of one argument $v(y)$ which we also consider to be the potential in Eq. (\ref{b0}). It is straightforward to verify that in this case the isometry group of the metric involves five Killing vectors
\bea\label{kvf}
&&
K_1=\partial_t, \qquad \qquad \quad K_2=\partial_s, \qquad \qquad \quad  K_3=\partial_x,
\nonumber\\[2pt]
&&
K_4=t \partial_x-y \partial_s, \qquad K_5=-y \partial_t+s \partial_x-2 \left(\int\limits_0^y v(r) d r \right) \partial_s,
\eea
which obey the algebra
\be
[K_1,K_4]=K_3, \qquad [K_2,K_5]=K_3.
\ee
$K_2$ and $K_3$ are covariantly constant and $K_2$, $K_3$, $K_4$ are null. The second rank Killing tensor associated with the quadratic integral of motion in (\ref{b0}) reads
\be
K_{tt}=-y v(y)+\int\limits_0^y v(r) dr, \qquad K_{xy}=-\frac{y}{2}, \qquad K_{yy}=x.
\ee
Together with $g_{AB}(z) \frac{d z^A}{d \tau} \frac{d z^B}{d \tau}$ they provide seven functionally independent integrals of motion which render the resulting geodesic flow maximally superintegrable.

As an example, let us consider the following system:
\be\label{Dr}
H=p_x p_y+\frac{a y}{\sqrt{y^2+b^2}}, \qquad I_1=p_x, \qquad I_2=p_x (x p_x-y p_y)+\frac{a b^2}{\sqrt{y^2+b^2}},
\ee
where $a$ and $b$ are arbitrary parameters, which is a particular case of the models considered by Drach \cite{Drach} in which a cubic integral of motion decomposes into a product of the linear and quadratic integrals of motion exposed in Eq. (\ref{Dr}). Within the Eisenhart framework, $I_1$ is related to the Killing vector field $K_3$ in (\ref{kvf}), while $I_2$ gives rise tho the second rank Killing tensor
\be
K_{tt}=\frac{a b^2}{\sqrt{y^2+b^2}}, \qquad K_{xy}=-\frac{y}{2}, \qquad K_{yy}=x.
\ee

The null geodesic equations (\ref{em}) and (\ref{first}) can be readily integrated. The leftmost equation in (\ref{em}) implies that $y(t)$ is a linear function of $t$
\be\label{1}
y(t)=y_0+y_1 t,
\ee
where $y_0$ and $y_1$ are constants of integration. If $y_1\ne 0$, the remaining equations in (\ref{em}) and (\ref{first}) yield\footnote{Recall that the constant $-c_2$ entering Eq. (\ref{first}) is interpreted as the value of the conserved energy. For the case at hand it reads $c_2=-x_1 y_1$. If $y_1=0$, one instead has $c_2=-\frac{a y_0 }{\sqrt{b^2 + y_0^2}}$. }
\bea\label{2}
&&
x(t)=-\frac{a\sqrt{b^2+y(t)^2}}{y_1^2}+x_0+x_1 t, \qquad s(t)=-x_1 y(t)+\frac{2 a \sqrt{b^2+y(t)^2}}{y_1}+s_0,
\eea
where $x_0$, $x_1$, $s_0$ are constants of integration.  Note that the same result could be attained by working directly with the seven integrals of motion mentioned above and algebraically expressing $t$, $s$, $x$, $y$ as
functions of $\tau$.

If the initial condition in (\ref{1}) is so chosen that $y_1=0$, then the solutions are simplified
\be\label{3}
y(t)=y_0, \qquad x(t)=-\frac{a {(b t)}^2}{2 {(b^2 + y_0^2)}^{\frac 32}}+x_0+x_1 t, \qquad s(t)=\frac{a y_0 t}{\sqrt{b^2 + y_0^2}}+s_0.
\ee
As is obvious from the form of these solutions, the spacetime is geodesically complete as geodesics emanating from a given point can be extended to infinite values of the affine parameter in both directions.

Taking into account the affine relations (\ref{afr}), (\ref{1}), it is easy to write down the solution to the geodesic equations for a generic continuously differentiable potential $v(y)$  
\begin{align}
&
t=c_1 \tau+t_0, && y(t)=y_0+y_1 t, 
\nonumber\\[2pt]
&
x(t)=-\frac{1}{y_1^2} \int\limits_0^{y(t)} v(r) dr+x_1 y(t)+x_0, && s(t)=\frac{2}{y_1} \int\limits_0^{y(t)} v(r) dr-x_1 y_1 y(t)+s_0,
\end{align}
provided $y_1\ne0$, while for $y_1=0$ one has
\be
t=c_1 \tau+t_0, \qquad y(t)=y_0, \qquad x(t)=-\frac{v'(y_0) t^2}{2}+x_1 t+x_0, \qquad s(t)=v(y_0)t+s_0. 
\ee
That the solutions are everywhere continuous functions guarantees that the spacetime is complete.

Concluding this section, we note that the additional constraint
\be
\lim\limits_{y\to \pm\infty}^{} v(y)=0
\ee
singles out an asymptotically flat spacetime.

\vspace{0.5cm}

\noindent
{\bf 4. Higher--dimensional generalizations}\\

\noindent
The self--dual metrics constructed in the preceding section can be used to generate solutions to the vacuum Einstein equations formulated in higher--dimensional spacetimes of the ultrahyperbolic signature $(2,q)$ with $q>2$. It suffices to consider a maximally superintegrable system of the type (\ref{b0}) whose potential $v(y)$ involves free parameters and apply to it the oxidation procedure. Given an integrable system, the oxidation yields another integrable system
with more degrees of freedom by converting each free parameter\footnote{To be more precise, free parameters should appear as overall factors.} entering the potential into a momentum conjugate to an extra cyclic variable. The standard kinetic term associated with the new canonical pair is to be
added to the Hamiltonian of the original mechanical system as well.

As an example, let us consider an asymptotically flat spacetime generated by the potential
\be
v(y)=g e^{-{\left(\frac{y}{m}\right)}^2},
\ee
where $g$ and $m$ are free parameters, and perform the oxidation with respect to $g$
\be\label{tH}
\tilde H=\frac 12 p_w^2+p_x p_y+p_w e^{-{\left(\frac{y}{m}\right)}^2}.
\ee
Here $(w,p_w)$ is a new canonical pair obeying the conventional Poisson bracket $\{w,p_w\}=1$. Because the variable $w$ is cyclic, the momentum $p_w$ is conserved. After the substitution $g\to p_w$, the function
$I_2$ in Eq. (\ref{b0}) gives the quadratic integral of motion. Note that, in general, to each extra canonical pair there corresponds only one new integral of motion which implies that the extended model will
lack one integral of motion to be maximally superintegrable.

Applying the Legendre transform with respect to the momenta $(p_x,p_y,p_w)$, one obtains the Lagrangian
\be
\mathcal{L}=\frac 12 {\dot w}^2+\dot x\dot y-\dot w e^{-{\left(\frac{y}{m}\right)}^2}+\frac 12 e^{-2{\left(\frac{y}{m}\right)}^2},
\ee
from which the Eisenhart metric is obtained
\be\label{gen}
d \tau^2=e^{-2{\left(\frac{y}{m}\right)}^2} dt^2+2 dt ds+2 dx dy-2 e^{-{\left(\frac{y}{m}\right)}^2} dt dw+dw^2.
\ee
Taking into account the relations which link the momenta and velocities
\be
\dot x=p_y, \qquad \dot y=p_x, \qquad \dot w=p_w+e^{-{\left(\frac{y}{m}\right)}^2},
\ee
one can readily compute components of the second rank Killing tensor of the metric (\ref{gen})
\begin{align}
&
K_{tt}=y e^{-2{\left(\frac{y}{m}\right)}^2}-\frac{m \sqrt{\pi}}{2} e^{-{\left(\frac{y}{m}\right)}^2} \mbox{erf} \left(\textstyle{\frac{y}{m}}\right), && K_{yy}=x,
\nonumber\\[2pt]
&
K_{tw}=-\frac{y}{2} e^{-{\left(\frac{y}{m}\right)}^2} +\frac{m \sqrt{\pi}}{4} \mbox{erf} \left(\textstyle{\frac{y}{m}}\right), && K_{xy}=-\frac{y}{2},
\end{align}
where $\mbox{erf} \left(y\right)=\frac{2}{\sqrt{\pi}} \int\limits_0^y e^{-r^2} dr$ is the error function.

If the potential $v(y)$ contains several terms, each of which involving a free parameter as an overall factor, one can implement the oxidation with respect to each parameter thus yielding a Ricci--flat metric of the ultrahyperbolic signature $(2,q)$ with $q>2$. In general, it will have the form
\be\label{final}
d \tau^2=a_0(y) dt^2+2 dt ds+2 dx dy+2 \sum\limits_{i=1}^m a_i(y) dt dw_i+ \sum\limits_{i=1}^m dw_i^2,
\ee
where $w_i$, $i=1,\dots,m$, denote extra coordinates and $a_i (y)$, $i=0,\dots,m$, are arbitrary continuously differentiable functions. As explained below Eq. (\ref{tH}), the geodesic flow associated with the metric (\ref{final}) is superintegrable but not maximally superintegrable. The derivation of a second rank Killing tensor proceeds along the same lines as in the five--dimensional example considered above.

\vspace{0.5cm}

\noindent
{\bf 5. Conclusion}\\

\noindent
To summarize, in this work we extended the recent study of spacetimes of the ultrahyperbolic signature $(2,q)$ with $q \geq 2$, which admit higher rank Killing tensors \cite{CG}, to the case of Killing tensors of valence two.
A class of self--dual metrics with maximally superintegrable geodesic flows has been constructed by applying the Eisenhart lift to special mechanics in pseudo--Euclidean spacetime of signature $(1,1)$.
As compared to \cite{CG}, the principal advantage of the spacetimes is that they are geodesically complete provided the potential governing the original mechanics is a continuously differentiable function. In dimension greater than four, our analysis yields superintegrable geodesic flows which, however, are not maximally superintegrable.

Turning to possible further developments, it would be interesting to study whether the solutions in this paper may find some applications within the context of the $N=2$ string theory (for a review see, e.g., Ref. \cite{BGPPR}).
Systematic extension of Drach's work \cite{Drach} to the case of quartic (or higher) integrals of motion is worthy of investigation as well. From a physical point of view, an important problem to study is
whether an integrable system with a cubic (or higher) integral of motion can be constructed in Euclidean space which is governed by a harmonic potential. This would allow one to build Ricci--flat spacetimes
of Lorentzian signature $(1,q)$ with $q>3$ admitting higher rank Killing tensors by applying the Eisenhart lift.
\vspace{0.5cm}

\noindent{\bf Acknowledgements}\\

\noindent
This work was supported by the MSE program Nauka under
the project 3.825.2014/K, the TPU grant LRU.FTI.123.2014, and the RFBR grant 15-52-05022. S.F. acknowledges the support of the Dynasty Foundation.

\end{document}